\documentclass[aip,graphicx]{revtex4-1}

\usepackage{t1enc}
\usepackage[latin1]{inputenc}
\usepackage{graphics}
\usepackage{graphicx}
\usepackage{amsmath,amsthm,amsfonts,amssymb,amsxtra,amsopn}
\usepackage{color}
\usepackage{sistyle}
\usepackage{exscale} 
\usepackage{epsfig} 

\begin{document}

\title{Evidence for topological band inversion of the phase change material Ge$_2$Sb$_2$Te$_5$} 

\author{Christian Pauly}
\affiliation{II.\ Inst.\ Phys.\ B and JARA-FIT, RWTH Aachen University, 52074 Aachen, Germany}

\author{Marcus Liebmann}
\affiliation{II.\ Inst.\ Phys.\ B and JARA-FIT, RWTH Aachen University, 52074 Aachen, Germany}

\author{Alessandro Giussani}
\affiliation{Paul-Drude-Institut f{\"u}r Festk{\"o}rperelektronik, Hausvogteiplatz 5-7, 10117 Berlin, Germany}

\author{Jens Kellner}
\affiliation{II.\ Inst.\ Phys.\ B and JARA-FIT, RWTH Aachen University, 52074 Aachen, Germany}

\author{Sven Just}
\affiliation{II.\ Inst.\ Phys.\ B and JARA-FIT, RWTH Aachen University, 52074 Aachen, Germany}

\author{Jaime S\'{a}nchez-Barriga}
\affiliation{Helmholtz-Zentrum f{\"u}r Materialien und Energie, Elektronenspeicherring BESSY II, Albert-Einstein-Strasse 15, 12489 Berlin, Germany}

\author{Emile Rienks}
\affiliation{Helmholtz-Zentrum f{\"u}r Materialien und Energie, Elektronenspeicherring BESSY II, Albert-Einstein-Strasse 15, 12489 Berlin, Germany}

\author{Oliver Rader}
\affiliation{Helmholtz-Zentrum f{\"u}r Materialien und Energie, Elektronenspeicherring BESSY II, Albert-Einstein-Strasse 15, 12489 Berlin, Germany}

\author{Raffaella Calarco}
\affiliation{Paul-Drude-Institut f{\"u}r Festk{\"o}rperelektronik, Hausvogteiplatz 5-7, 10117 Berlin, Germany}

\author{Gustav Bihlmayer}
\affiliation{Peter Gr{\"u}nberg Institut (PGI-1) and Institute for Advanced Simulation (IAS-1), Forschungszentrum J{\"u}lich and JARA, 52425 J{\"u}lich, Germany}

\author{Markus Morgenstern}
\affiliation{II.\ Inst.\ Phys.\ B and JARA-FIT, RWTH Aachen University, 52074 Aachen, Germany}


\date{\today}

\begin{abstract}
We present an angle-resolved photoemission study of a ternary phase change material, namely Ge$_2$Sb$_2$Te$_5$, epitaxially grown on Si(111) in the metastable cubic phase. The observed upper bulk valence band shows a minimum at $\bar{\Gamma}$ being 0.3\,eV below the Fermi level $E_{\rm F}$ and a circular Fermi contour around $\bar{\Gamma}$ with a dispersing diameter of $0.27-0.36$\,{\AA}$^{-1}$. This is in agreement with density functional theory calculations of the Petrov stacking sequence in the cubic phase which exhibits a topological surface state. The topologically trivial cubic KH stacking shows a valence band maximum at $\Gamma$ in line with all previous calculations of the hexagonal stable phase exhibiting the valence band maximum at $\Gamma$ for a trivial ${\mathbb Z}_2$ topological invariant $\nu_0$ and away from $\Gamma$ for non-trivial $\nu_0$. Scanning tunneling spectroscopy exhibits a band gap of 0.4 eV around $E_{\rm F}$.
\end{abstract}


\maketitle 

\section{Introduction}

Following the proposal\cite{volkov85,fu07} and discovery\cite{koenig07,hsieh08} of topological insulators (TIs), materials are currently optimized in terms of separating the Dirac cone from bulk bands and tuning the Dirac point close to the Fermi energy $E_{\rm F}$. In this course, compounds involving more than two elements are preferentially used since they offer more degrees of freedom.\cite{eremeev12,ando13} Connecting such compounds to classes of materials already in use for electronic or storage applications is desirable towards the utilization of topological properties. An important material system for commercially used optical and non-volatile electrical data storage are phase change materials (PCMs),\cite{wuttig07,wuttig12} which are found predominantly along the pseudobinary line connecting GeTe and Sb$_2$Te$_3$.\cite{lencer08}

Such PCMs exhibit a large contrast in electronic and optical properties upon changing from amorphous to crystalline.\cite{ovshinsky68,yamada87} Using laser-induced or electrical heat pulses, the switching occurs within nanoseconds\cite{yamada91} or below\cite{loke12} at an energy cost of only 1\,fJ.\cite{xiong11}
The PCM Sb$_2$Te$_3$ is experimentally known to be a TI \cite{hsieh09a,zhang09b,pauly12} and some of the other compounds on the pseudobinary line are predicted to be TIs based on density functional theory (DFT) calculations.\cite{kim10,eremeev12,kim12,kim12a,silkin13} Ge$_2$Sb$_2$Te$_5$ (GST-225) is at the borderline of these predictions,\cite{kim12,silkin13} i.e., its TI properties depend on the stacking sequence.\cite{silkin13} Here, we present experimental evidence for the non-trivial topology of GST-225 by angle-resolved photoemission spectroscopy (ARPES), supported by DFT calculations. The result implies that half of the pseudobinary line consists of TIs and opens up the perspective for fast and reversible switching between a crystalline topological phase and an insulating amorphous phase.

GST-225, a prototype PCM, emerges in two slightly different crystalline phases, a metastable cubic one used for applications\cite{park07}
and a stable hexagonal one. Within the stable phase, hexagonal layers are stacked along [0001] with a sequence deduced from transmission electron microscopy (TEM) to be either Te-Sb-Te-Ge-Te-v-Te-Ge-Te-Sb- (Petrov phase)\cite{petrov68} or Te-Ge-Te-Sb-Te-v-Te-Sb-Te-Ge- (Kooi-De Hosson or KH phase).\cite{kooi02} The v denotes a vacancy layer where adjacent Te layers are van-der-Waals bonded. DFT calculations imply that the KH phase is energetically favorable.\cite{sun06} More recent X-ray diffraction data suggest some mixture of Ge and Sb in the respective layers.\cite{matsunaga04}
The cubic rocksalt structure exhibits hexagonal layers stacked along [111] with (Te$-$Ge/Sb/v)$_3$ sequence, where Ge/Sb/v is a mixed layer of Ge, Sb and vacancies.\cite{yamada00, matsunaga04} More recent TEM studies suggest that the Ge/Sb/v layers exhibit some internal order\cite{park05} and DFT even implies that Ge, Sb and vacancies accumulate in separate layers.\cite{sun06} Thus, the stable and the metastable phase could be much closer than originally anticipated. Then, the transition between them would be a mere shift of blocks of (111) layers without atomic rearrangements within the layers.\cite{sun06}

The first prediction of topologically insulating GST-225 was made by Kim {\it et al}.\ for the Petrov phase while the energetically favorable KH phase was shown to be topologically trivial.\cite{kim10} However, even the KH phase of GST-225 can be made a TI by DFT if set under isotropic pressure\cite{sa11} or strain.\cite{sa12} A more disordered, hexagonal mixed-layer phase was investigated by Silkin {\it et al}.\ by DFT with the stacking sequence Te-M1-Te-M2-Te-v-Te-M2-Te-M1- having (Ge$_{2x}$Sb$_{2(1-x)}$ in M1 and Ge$_{2(1-x)}$Sb$_{2x}$ in M2).\cite{silkin13} The transition between the Petrov ($x=0$) and the KH ($x=1$) phase exhibits a semimetal for the Petrov phase, a trivial band insulator for $x=1$ and $x=0.75$, and a topological insulator for $x=0.25$ and $x=0.5$.

Importantly, all DFT calculations of GST-225 exhibiting the valence band maximum (VBM) away from $\Gamma$ show topologically non-trivial properties.\cite{kim10,kim12,sa11,sa12,silkin13} The only exception is the Petrov phase calculated by Silkin {\it et al}.\ which is semimetallic.\cite{silkin13} This empirical relation is our central argument in the following.

So far, there have been no calculations including spin-orbit coupling for the metastable rocksalt phase. We will provide them for the Petrov and KH stacking, confirming the above trend. To ease the comparison of our data also with previous calculations, we stick to the hexagonal nomenclature also for the metastable cubic phase, identifying the cubic [111] with the hexagonal [0001] direction.

In order to study TI properties by angle-resolved photoelectron spectroscopy (ARPES), ideally single crystalline GST is desired. Typically, GST is deposited in a polycrystalline fashion by physical vapor deposition. Only recently, epitaxial films of superior crystalline quality have been grown molecular beam epitaxy (MBE) on GaSb, InAs, and Si.\cite{katmis11,rodenbach12,takagaki12,giussani12,takagaki13} The metastable cubic, rhombohedrally distorted GST-225 grows with a single vertical epitaxial orientation, well defined interfaces, and atomically flat terraces only on (111)-oriented substrates.\cite{katmis11,rodenbach12}

\section{Preparation}

The GST-225 layers were grown on Si(111) in an MBE machine dedicated to chalcogenides.\cite{supp} The temperature of the effusion cells was set such that the flux ratio of Ge:Sb:Te is close to 2:2:5, as confirmed by X-ray fluorescence.\cite{katmis11,rodenbach12} Out-of-plane and grazing incidence X-ray diffraction (XRD) shows that GST grows in the metastable cubic phase along the [111] direction. The presence of superstructure peaks in addition to the Bragg reflections of cubic  GST-225 indicates vacancy ordering in the Ge/Sb/v sublattice along the growth direction. The film thickness was 20\,nm, and the growth temperature 250\,°C.

After growth, the samples have been transferred under ambient conditions. Before insertion into the ultrahigh vacuum ARPES or scanning tunneling microscopy (STM) chambers, the surface was deoxidized by dipping in de-ionized (DI) water for one minute following the procedure of Zhang {\it et al}.\cite{zhang10} Afterwards, the sample was introduced within two minutes and, after pump-down, annealed to 250\,°C yielding a clean crystalline, stoichiometric and oxygen-free surface.\cite{supp} XRD data confirm that neither this procedure nor the subsequent measurements change the phase of the GST-225.

\begin{figure}
\includegraphics[width=8.5cm]{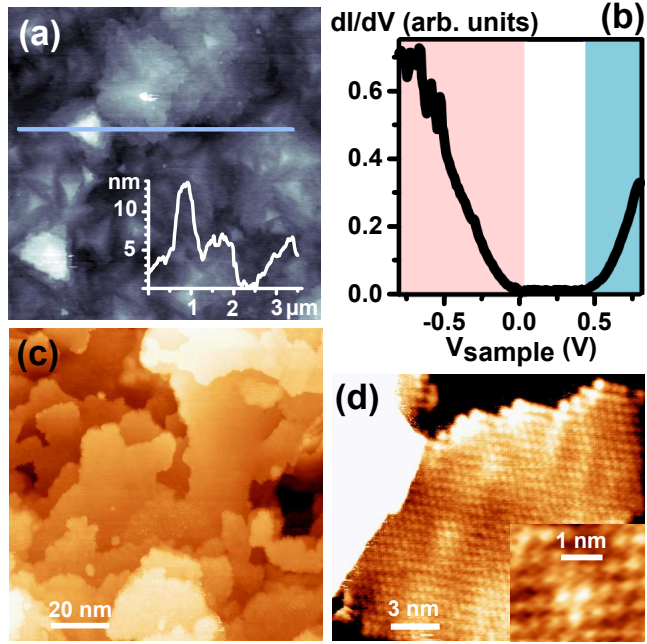}%
\caption{\label{fig1}Scanning probe microscopy of metastable GST-225 (111) after DI water dip. (a) Tapping mode AFM under ambient conditions. Inset: profile along the line marked in the image. (b) STS curve recorded in UHV (average of 10 spectra). Red/light shaded (blue/darker) area marks the valence (conduction) band. Stabilization at $V_\mathrm{sample}=-0.8$\,V, $I=100$\,pA. (c) STM image of atomically flat terraces. Average step height: 0.34\,nm. $V_\mathrm{sample}=-0.3$\,V, $I=100$\,pA. (d) STM image with atomic resolution (inset: zoom). $V_\mathrm{sample}=-0.5$\,V, $I=100$\,pA. All data are taken at room temperature.}
\end{figure}

The topography was investigated by atomic force microscopy (AFM) on a $\mu$m scale and by STM on the nm scale.
The AFM topography (Fig.\ \ref{fig1}(a)) exhibits an overall roughness of 3-4\,nm$_\mathrm{rms}$ due to pyramids with $5-15$\,nm in height and a width close to 1\,$\mu$m. On their slopes, atomically flat terraces up to 100\,nm in width are found (Fig.\ \ref{fig1}(c)).
These terraces are separated by steps of $0.34 \pm 0.01$\,nm in height, in agreement with the expected Te-Te layer distance of 0.347\,nm in cubic GST-225.\cite{nonaka00} On the terraces, atomic resolution is achieved by STM (Fig.\ \ref{fig1}(d)), most likely showing the Te layer.\cite{deringer13} STS shows a band gap of 0.4\,eV with $E_{\rm F}$ situated at the top of the valence band (Fig.\ \ref{fig1}(b)).

\section{Results}

\begin{figure}
\includegraphics[width=8.5cm]{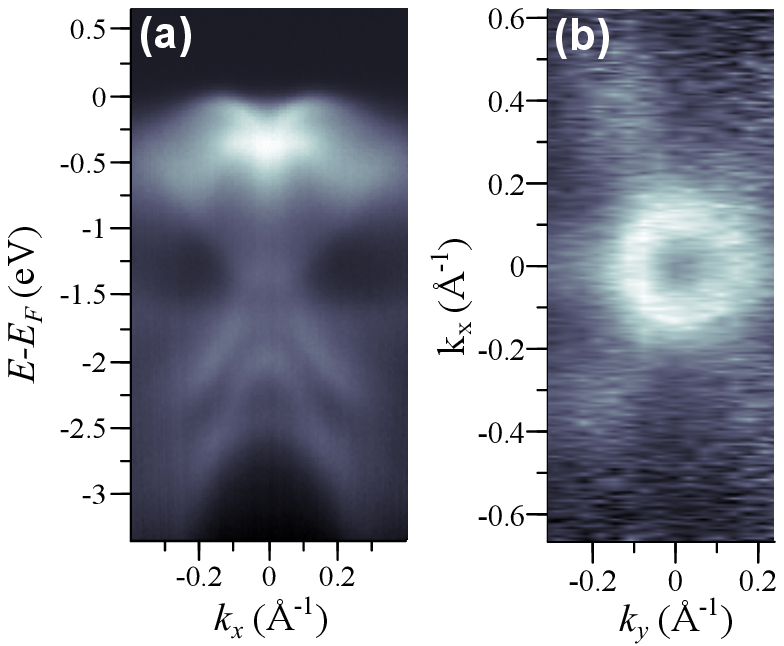}%
\caption{\label{fig2}ARPES spectra of metastable GST-225 (111) after DI water dip: (a) recorded in $\bar{\Gamma}-\bar{\mathrm{K}}$ direction (see Ref.\ \onlinecite{supp}, Shirley-type background subtracted), (b) Intensity in the $k_\mathrm{||}$ plane at $E_{\rm F}$, $\bar{\Gamma}-\bar{\mathrm{M}}$ direction is horizontal; photon energy: 22\,eV, temperature: 300\,K.}
\end{figure}

ARPES measurements have been performed at the beamline UE112-lowE-PGM2 ($1^2$) at BESSY II using a Scienta R8000 analyzer. Figure \ref{fig2}(a) displays a spectrum recorded with linearly polarized light at $h\nu = 22$\,eV in a direction determined to be $\bar{\Gamma}-\bar{\mathrm{K}}$ by comparison with DFT calculations.\cite{supp} Just below $E_{\rm F}$, the upper valence band shows maxima at $k_\mathrm{||} = \pm 0.14 \pm 0.02$\,{\AA}$^{-1}$ and drops to $E - E_\mathrm{F} = -0.3$\,eV at $\bar{\Gamma}$. Another band resides between -0.7\,eV at $k_\mathrm{||} = \pm 0.23$\,{\AA}$^{-1}$ and $-0.35$\,eV at $k_\mathrm{||} = \pm 0.1$\,{\AA}$^{-1}$. Closer to $\bar{\Gamma}$, these two bands lead to a broad peak in energy distribution curves (EDCs) around $-0.4$\,eV with a FWHM of $0.5$\,eV (Fig.\ \ref{fig4}(a)). Below $-1$\,eV, there are two more hole-like bands.
The $\bar{\Gamma}-\bar{\mathrm{M}}$ direction looks essentially the same with slightly more intensity at even higher $|k|$ values.\cite{supp} This can be seen from the constant energy cut at $E_{\rm F}$  in Fig.\ \ref{fig2}(b), showing a nearly isotropic circle and faint additional intensity at high $|k|$ values in the six different $\bar{\Gamma}-\bar{\mathrm{M}}$ directions.

\begin{figure}
\includegraphics[width=8.5cm]{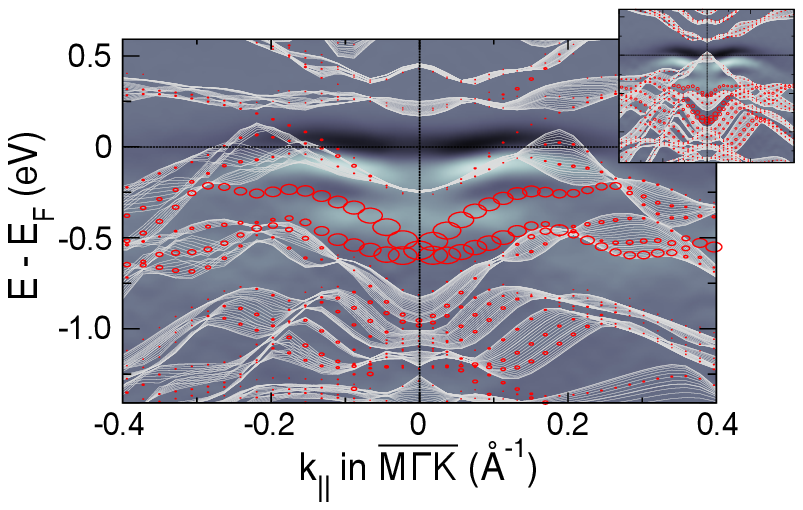}%
\caption{\label{fig3}DFT calculations of the band structures for cubic GST-225 with Petrov- and KH-type (inset, same scale as main image) stacking sequence, as proposed in Ref.\ \onlinecite{sun06}. Bulk bands are given as gray lines, states of the film calculations with circles. The extension of the states into the vacuum (region above the topmost Te layer) is indicated by the size of the circles. The calculations are superimposed with the ARPES spectra (2$^\mathrm{nd}$ derivative of intensity with respect to electron energy) at 22\,eV photon energy. Calculations are shifted upwards by 100\,meV.}
\end{figure}

DFT calculations have been performed within in the generalized gradient approximation.\cite{perdew96} We employed the full-potential linearized augmented planewave method in bulk and thin-film geometry \cite{krakauer79} as implemented in the FLEUR code.\cite{supp} Spin-orbit coupling was included self-consistently and a basis set cutoff of $R_{\rm MT} k_{\rm max} = 9$ was used. As structural model for the cubic phases we adopted the atomic positions given by Sun {\itshape et al.},\cite{sun06}
both for the bulk and film structures. For the latter, films consisting of 27 atomic layers terminating by a 'vacancy layer' were used. Two different stacking sequences were assumed for the cubic phase: a Petrov- and a KH-like sequence which are derived from the respective hexagonal phases by tripling the unit cell and adding appropriate shifts.

Figure \ref{fig3} shows the $2^\mathrm{nd}$ derivative of the measured band structure along with the calculations. A reasonable agreement is only achieved with the Petrov-like stacking, including the minimum at $\bar{\Gamma}$ of the upper valence band. The bands further down in energy (around $-0.6$\,eV at $\bar{\Gamma}$) can be associated with a Rashba-type surface state, similar to the one observed in Sb$_2$Te$_3$.\cite{pauly12} In close vicinity of the upper valence band, the calculation shows the topological surface state crossing the Fermi energy at $k_{||} \approx 0.12$\,{\AA}$^{-1}$. This state obviously overlaps with the upper bulk valence band within our ARPES data.

\begin{figure}
\includegraphics[width=8.5cm]{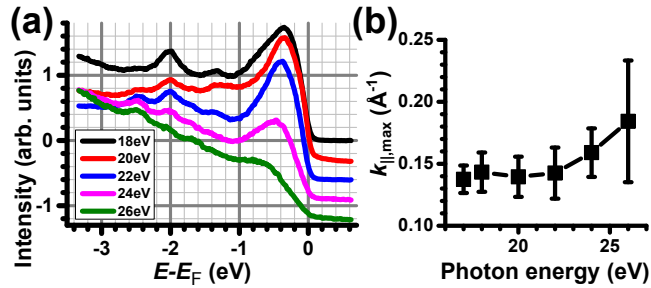}%
\caption{\label{fig4}Effects of photon energy variation. (a) EDCs at $\bar{\Gamma}$ for photon energies $h\nu=18-26$\,eV (top-down) as indicated, corresponding to a $k_z$ variation of $\Delta k_z \approx 0.35$\,{\AA}$^{-1}$. Graphs are offset for clarity. (b) $k_\mathrm{||}$-value of the valence band maximum ($\bar{\Gamma} - \bar{K}$ direction) as extracted from spectra at different photon energies.}
\end{figure}

In order to probe different $k_z$ and, thus, distinguish surface from bulk states, spectra with $h\nu = 17-26$\,eV have been recorded.\cite{supp} Corresponding EDCs at $\bar{\Gamma}$ are shown in Fig.\ \ref{fig4}(a). The topmost maximum shifts down by about 0.2\,eV between $h\nu=22$\,eV and 26\,eV, indicating a $k_z$ dispersion, as expected for a band with bulk character. The $k_\mathrm{||}$-position of the VBM has been evaluated by means of the 2$^\mathrm{nd}$ derivative of the ARPES spectra: the EDC with the peak at the highest energy is chosen, and the corresponding $k_\mathrm{||}$ value is taken as $k_\mathrm{||,max}$.\cite{supp} Figure \ref{fig4}(b) displays $k_\mathrm{||,max}$ as a function of photon energy revealing a small dependence on $k_z$ as well. Thus, the ARPES peak at the VBM is, at least partially, a bulk band with dispersion in $k_z$ direction.

\section{Discussion}

The calculated $k_{||,\mathrm{max}}$ of the bulk valence band ($0.19-0.22$\,{\AA}$^{-1}$) is larger than the experimental one ($0.14-0.18$\,{\AA}$^{-1}$). This can be explained by the overlap with the surface state which crosses $E_\mathrm{F}$ at $k_{||} \approx 0.12$\,{\AA}$^{-1}$ (Fig.\ \ref{fig3}). The small anisotropy of $k_{||,\mathrm{max}}$ (DFT: 7\%, ARPES: <10\%) between $\bar{\Gamma}-\bar{M}$ and $\bar{\Gamma}-\bar{K}$ was not detectable within the experimental error.

We finally compare the metastable cubic phase with previous DFT calculations of the very similar hexagonal phase. Most notably, a VBM away from $\bar{\Gamma}$ consistently indicates topologically non-trivial properties for GST-225.\cite{kim10,kim12,sa11,sa12,silkin13} Albeit such a relation is also found for the 3D TIs BiSb\cite{hsieh08,hsieh08a}, Bi$_2$Te$_3$\cite{hsieh09,zhang09b}, and Sb$_2$Te$_3$\cite{hsieh09a,zhang09b,pauly12}, it is currently under discussion for Bi$_2$Se$_3$.\cite{xia09,zhang09b,nechaev13} None of these materials have a VBM away from $\Gamma$ with trivial properties.\cite{aguilera13} The measured $k_\mathrm{||,max}$ is smaller than the calculated $k_\mathrm{||,max}$ of the bulk VBM of topologically non-trivial hexagonal stable phases of GST-225 ($0.16-0.52$\,{\AA}$^{-1}$).\cite{supp} The superposition of bulk valence band and topological surface state in the ARPES data might be relevant again.

\section{Summary}

In summary, we have shown by ARPES and STS that metastable cubic Ge$_2$Sb$_2$Te$_5$ epitaxially grown on Si(111) exhibits valence band maxima $0.14-0.18$\,{\AA}$^{-1}$ away from $\bar{\Gamma}$ and a band gap of 0.4 eV. All DFT calculations of Ge$_2$Sb$_2$Te$_5$ find a VBM away from $\Gamma$ only for a ${\mathbb Z}_2$ topological invariant $\nu_0=1$. This implies topological properties of Ge$_2$Sb$_2$Te$_5$, indicates that all phase change materials on the pseudobinary line between Sb$_2$Te$_3$ and Ge$_2$Sb$_2$Te$_5$ are topologically non-trivial, and opens up the possibility of switching between an insulating amorphous phase and a topological phase on ns time scales.

\begin{acknowledgments}
We gratefully acknowledge financial support by the German Science Foundation (DFG): SFB 917, project A3, the SPP 1666 ``Topological Insulators'' and Mo058/13-1; the Helmholtz-Zentrum Berlin (HZB); the excellence initiative of the German federal government; the Fonds National de la Recherche (Luxembourg). We thank F.\ Lange, S.\ Behnke and C.\ Stemmler for technical support.
\end{acknowledgments}

\section{Supplemental Information}

\subsection{Sample Preparation}

\begin{figure}[tb]
\includegraphics[width=16cm]{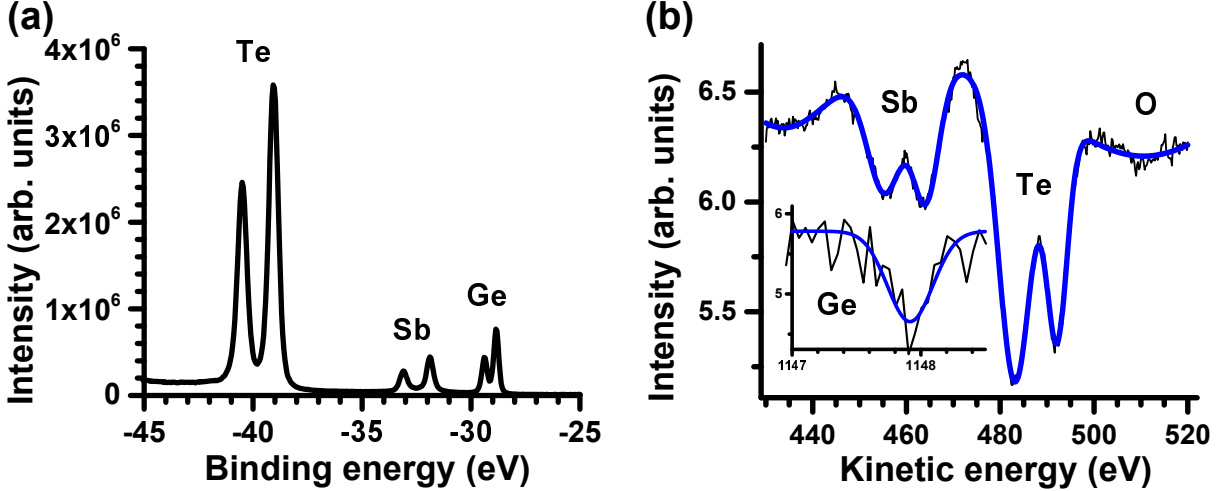}%
\caption{\label{supp1} GST-225 samples after de-ionized water dip and UHV annealing. (a) Photoelectron spectroscopy of the Ge 3d and the Sb 4d and Te 4d levels. Photon energy: 110\,eV. (b) Auger electron spectroscopy covering the peak positions of Ge, Sb, Te, and O as marked (black); the blue lines are Gaussian fits used to detemine the stoichiometry to be: Ge: 18\%, Sb: 21\%, Te: 57\%, O: 4\%.}
\end{figure}

The MBE system, located at the PHARAO beamline at BESSY II (Helmholtz Center for Materials and Energy, Berlin), is equipped with separate dual filament hot lip effusion cells for the evaporation of elemental Ge, Sb, and Te. In-situ reflection high-energy electron diffraction (RHEED), line-of-sight quadrupole mass spectrometry (QMS) and {\it in-situ} X-ray diffraction (XRD) by a six-circle diffractometer for synchrotron radiation were used to optimize the growth conditions.
After transport under ambient conditions, de-ionized (DI) water dip and annealing in UHV to 250\,°C, X-ray photoelectron spectroscopy has been routinely performed in the ARPES chamber. The observed peaks in Fig.\ \ref{supp1}(a) arise from the Ge 3d, Sb 4d and Te 4d levels.\cite{bearden67b,fuggle80} The absence of any distortion of the peaks implies a clean and oxygen-free surface.
The crystal structure and stoichiometry have been checked after the STM measurements by low-energy electron diffraction (LEED) and Auger electron spectroscopy (AES) (Fig.\ \ref{supp1}(b)), again after DI water dip and annealing in UHV. The stoichiometry has been calculated\cite{mroczkowski85} by using the tabulated sensitivities for Ge, Sb, Te, and O peaks. The larger than stoichiometric Te content of 57\% compared to Ge (18\%) and Sb (21\%) is partly attributed to the Te termination leading to larger AES intensity. Most notably, the oxygen content of the surface is only 4\%.

Further heating in vacuum above 300\,°C in order to induce a transition into the stable phase at ~340\,°C,\cite{kooi04} was found to cause considerable change in stoichiometry due to the different desorption temperatures of atomic species. At the same time, a change in the peak structures of the Ge 3d and Sb and Te 4d levels in photoelectron spectroscopy was observed.
We checked that neither the measurement processes nor our preparation lead to a phase transition. For that purpose, X-ray diffraction after the ARPES measurements was performed showing the same cubic structure with vacancy ordering.
Indeed, the power of the incident light of $10^{13}$ photons/s at $h\nu<26$\,eV on a spot area of $A=0.015$\,mm$^2$ can be estimated to be below 3,000\,W/m$^2$. Assuming the thermal conductivity of the Si substrate of 150 W/(m$\cdot$K) across the area $A$, a temperature gradient of just 20\,K/m, would be enough in order to create a sufficient heat flux. This value is low enough not to induce unintentional heating, or even a phase change of the material.

\subsection{Evaluation of additional ARPES spectra}

\begin{figure}[tb]
\includegraphics[width=16cm]{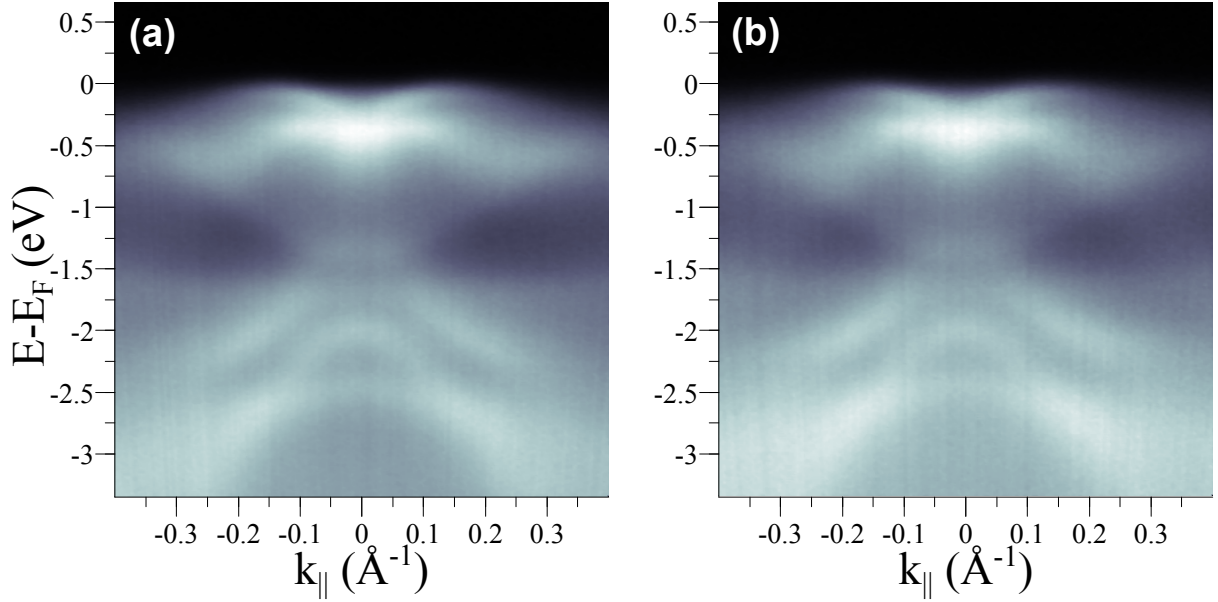}%
\caption{\label{supp2} ARPES spectra with 30° relative azimuth rotation, taken at 22\,eV photon energy. (a) $\bar{\Gamma}-\bar{K}$, (b) $\bar{\Gamma}-\bar{M}$ direction.}
\end{figure}

\begin{figure}[tb]
\includegraphics[width=16cm]{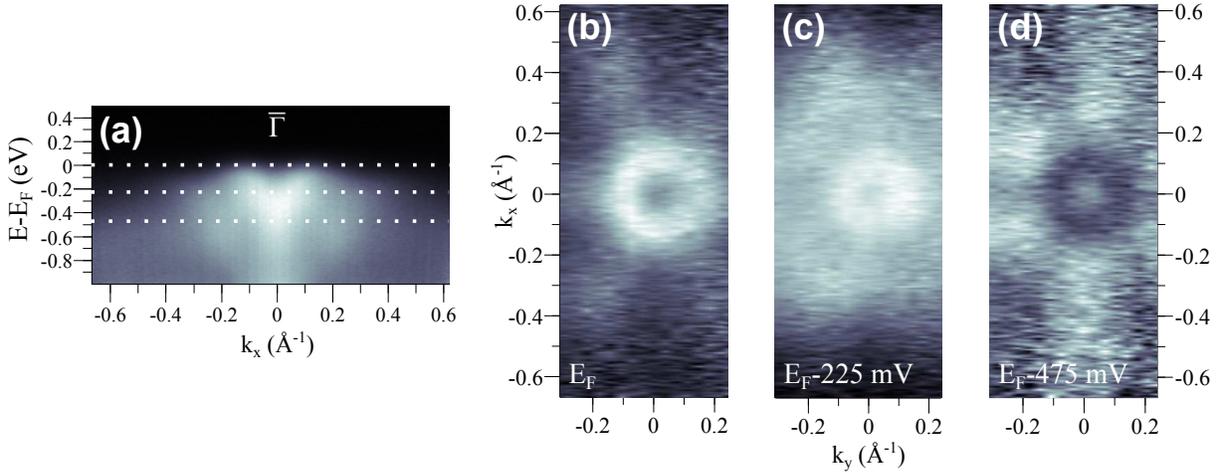}%
\caption{\label{supp3} (a) ARPES band structure taken in $\bar{\Gamma} - \bar{K}$ direction. Dotted lines mark the energies of the constant energy cuts in (b)-(d). (b)-(d) constant energy cuts in $k_\mathrm{||}$-directions at energies as indicated. At all energies, a sixfold symmetry is visible. Photon energy: 22\,eV.}
\end{figure}

Figure \ref{supp2} displays two spectra acquired at 22\,eV photon energy with 30° relative azimuth rotation in the sample plane. Between them, no difference can be observed at first sight. However, the plots at $E_\mathrm{F}$ (Fig.\ \ref{supp3}(b)-(d)) show a sixfold symmetry at larger $k_\mathrm{||}$ values, which persists down to, at least, -220\,meV. Since DFT calculations of cubic Ge$_2$Sb$_2$Te$_5$ do not show any bands at higher $k$ than $0.3$\,{\AA}$^{-1}$ down to $-0.2$\,eV in $\Gamma -K$ direction, but bands at such high $k$ values in $\Gamma - M$ (see Fig.\ \ref{supp7}(b)), we attribute the direction with intensity at high $k$ values in Fig.\ \ref{supp3}(b) to $\bar{\Gamma} - \bar{M}$. Thus, we denote Fig.\ \ref{supp2}(a) as $\bar{\Gamma}-\bar{K}$ and Fig.\ \ref{supp2}(b) as $\bar{\Gamma}-\bar{M}$. The same assignment is also used in the main text.

\begin{figure}[tb]
\includegraphics[width=16cm]{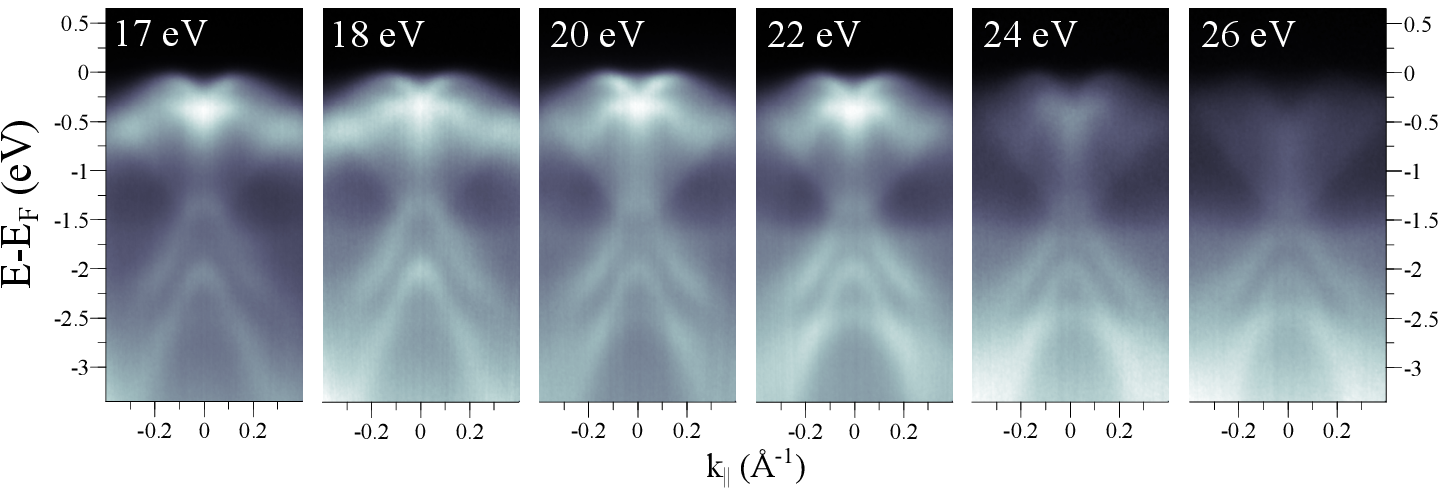}%
\caption{\label{supp4} ARPES band structure taken in $\bar{\Gamma} - \bar{K}$ direction at different photon energies as indicated.}
\end{figure}

\begin{figure}[tb]
\includegraphics[width=16cm]{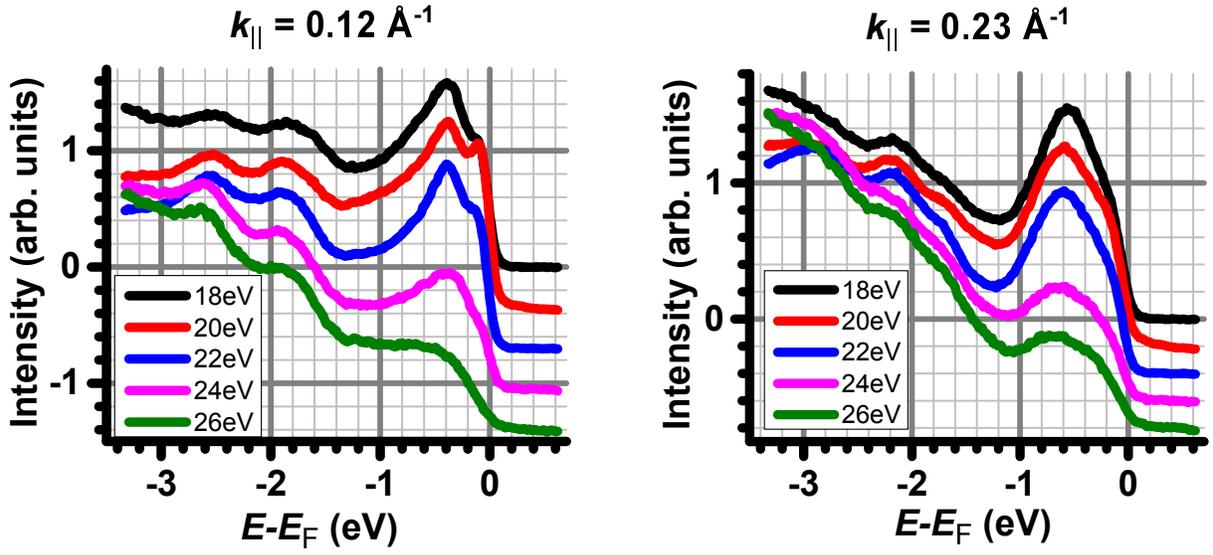}
\caption{\label{supp5} Energy distribution curves for different photon energies at two different $k_\mathrm{||}$ as indicated ($\bar{\Gamma} - \bar{K}$ direction). The set on the left is acquired near the position of the valence band maximum. The topmost band being separated from the much broader lower band can be identified. The maxima below -1\,eV mark the two states which do not disperse in the surface normal direction.}
\end{figure}

The effect of different photon energies is shown in Fig.\ \ref{supp4}. The upper valence band changes with photon energy revealing itself as a bulk band while the two bands below $-1$\,eV do not. Also in energy distribution curves (Fig.\ \ref{supp5}), no significant shift of these bands can be observed.

Although the film is in the cubic phase, as deduced from XRD data,\cite{rodenbach12} cuts at different energies as displayed in Fig.\ \ref{supp3} show a sixfold symmetry instead of a threefold one. The most probable reason is twinning, i.e., a 180° rotation of crystallographic domains around the surface normal.\cite{rodenbach12} Additionally, it is known from azimuthal XRD scans that the film exhibits rotational domains with a modulated angular spread of $10-15$°, a behavior typical for epitaxial films of GST,\cite{giussani12} Sb$_2$Te$_3$, and GeTe\cite{rodenbach12a} on Si(111). This twist leads to an angular broadening of the spectra and weakens the measured anisotropy between $\bar{\Gamma} - \bar{K}$ and $\bar{\Gamma} - \bar{M}$ directions, thereby making a better agreement of an untwinned film with the calculations of the bulk VBM more likely.

The energy distribution curves in Fig.\ \ref{supp5} show the development of the upper valence band at higher $k$ with photon energy, again indicating the bulk-like character by the observable dispersion.

\subsection{Determination and comparison of valence band maxima}

\begin{figure}[tb]
\includegraphics[width=16cm]{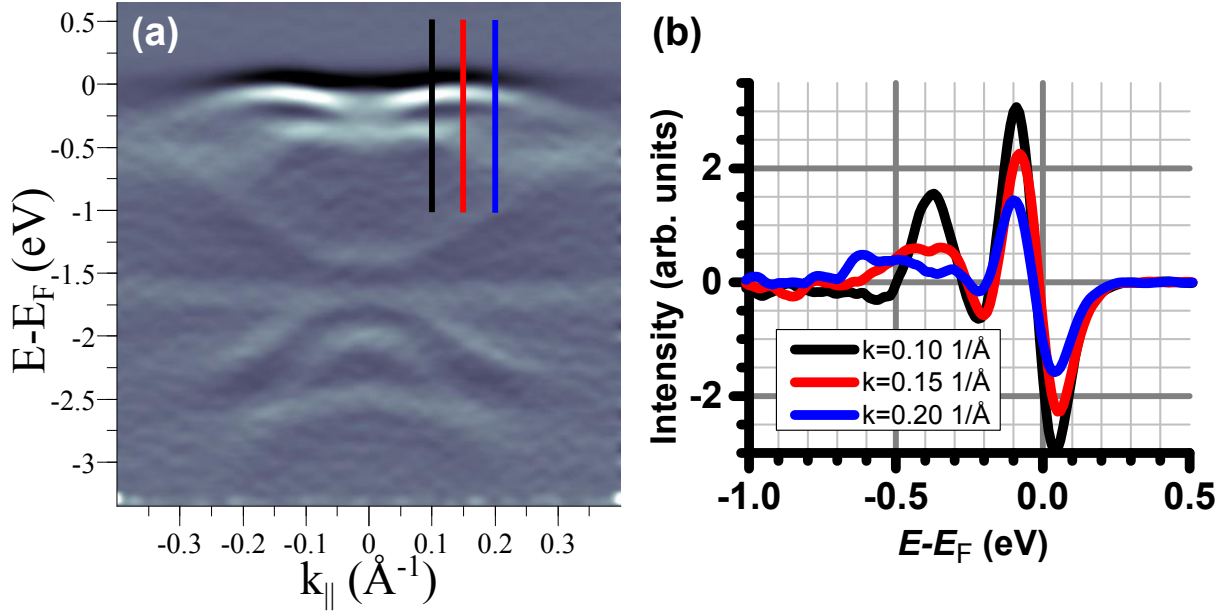}
\caption{\label{supp6} Determination of the valence band maxima. (a) Second derivative (ARPES intensity with respect to electron energy) of Fig.\ \ref{supp2}(a). (b) Cuts through the data (a) at exemplary constant $k_{||}$ values as marked. The $k_{||}$ value of the curve exhibiting the highest peak energy (the red one, in this case) is taken as $k_{||,\mathrm{max}}$ for this particular photon energy.}
\end{figure}

The maximum of the valence band has been determined for different photon energies by means of the second derivative of the ARPES intensity with respect to electron energy. Figure \ref{supp6} shows an example of the evaluation procedure. After computing the second derivative (Fig.\ \ref{supp6}(a)), energy distribution curves (EDCs) of these data along constant $k_{||}$ are extracted. The peak being closest to the Fermi energy marks the energy of the valence band at the given $k_{||}$. For determination of the $k_{||}$ value of the valence band maximum (VBM), EDCs for different $k_{||}$ are evaluated and the $k_{||}$ where the valence band peak is highest in energy is taken as the position of the VBM, $k_{||,\mathrm{max}}$. This procedure is applied for the ARPES spectra of 6 different photon energies (Fig.\ \ref{supp4}) and entered into Fig.\ 4(b) of the main text. Since only EDCs at constant $k_{||}$ are used, variations in ARPES intensity with detection angle, or $k_{||}$, do not influence the outcome.

\begin{table}
\caption{\label{tab1} $k_\mathrm{||,max}$ positions of experimental and theoretical valence band maxima given in {\AA}$^{-1}$, theoretical values from the literature are extracted from graphs in the cited publications using only the topologically non-trivial phases. The percentages (25 \%, 50 \%) denote the fraction of Ge in the M1 layer.}
\begin{ruledtabular}
\begin{tabular}{ccccccc}
\multicolumn{2}{c}{this work} & \multicolumn{2}{c}{Kim {\it et al.}\footnotemark[1]} & \multicolumn{3}{c}{Silkin {\it et al.}\footnotemark[2]} \\ \hline
 & cubic & & Petrov &  & 25\% & 50\% \\ \hline
$\Gamma - K$ (DFT) & 0.19 & $\Gamma - K$ & 0.18 & $\Gamma - K$ & 0.29 & 0.30 \\
$\Gamma - M$ (DFT) & 0.22 & $\Gamma - M$ & 0.26 & $\Gamma - M$ & 0.51 & 0.52 \\
$h \nu = 20$\,eV (exp., $\bar{\Gamma} - \bar{K}$) & 0.14 & $\bar{\Gamma}-\bar{K}$ & 0.18 & $A-H$ & 0.20 & 0.16 \\
$h \nu = 26$\,eV (exp., $\bar{\Gamma} - \bar{K}$) & 0.18 & $\bar{\Gamma}-\bar{M}$ & 0.21 & $A-L$ & 0.25 & 0.21 \\
\end{tabular}
\end{ruledtabular}
\footnotetext[1]{see Ref.\ \onlinecite{kim10}}
\footnotetext[2]{see Ref.\ \onlinecite{silkin13}}
\end{table}

Table \ref{tab1} compares the experimental values of the VBM to calculations of the hexagonal stable phase and the cubic metastable phase. Best agreement is found with the slab calculation of the Petrov phase\cite{kim10} and with the mixed phase with equal distribution of Ge and Sb ($x=0.5$).\cite{silkin13} Within the Brillouin zone of this phase, $k_{||,\mathrm{max}}$, the valence band maximum projected onto the (0001) plane, is closest to $\bar{\Gamma}$ at the edge of the Brillouin zone (see cuts connecting the $H-A-L$ points in Ref.\ \onlinecite{silkin13}).

\subsection{Density functional theory calculations of the metastable phase}

\begin{figure}[tb]
\includegraphics[width=16cm]{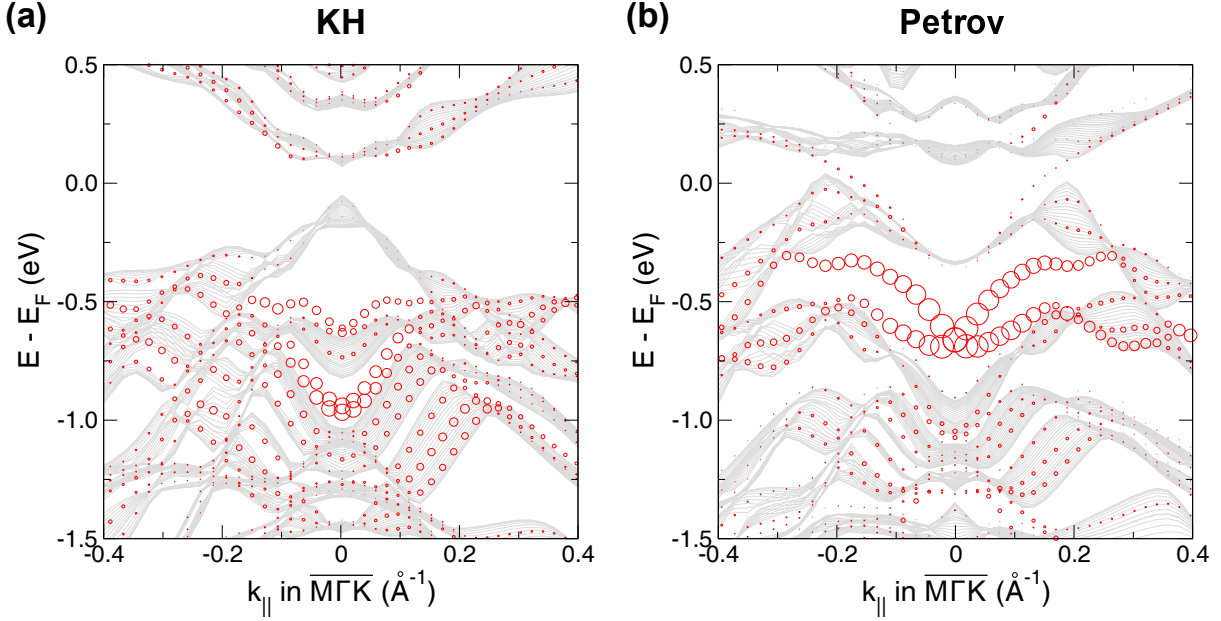}%
\caption{\label{supp7} Density functional theory calculations of the cubic metastable phase of GST-225. The structural model is taken from Sun {\it et al.} (Ref.\ \onlinecite{sun06}). Two stacking sequences have been assumed: a KH- and a Petrov-type stacking as marked. Only the Petrov-type stacking shows a valence band maximum away from $\Gamma$ and and a topological surface state which crosses the Fermi energy at $k_\mathrm{||} \approx 0.15$\,{\AA}$^{-1}$. Bulk bands are given as gray lines, states of the film calculations with circles. The extension of the states into the vacuum (region above the topmost Te layer) is indicated by the size of the circles.}
\end{figure}

Density functional theory calculations including spin-orbit coupling have been employed using the FLEUR code\cite{flapw} for the metastable cubic phase of GST-225. The structure has been derived from the hexagonal KH and Petrov phase by introducing a shift of one part of the unit cell within the [0001] plane, as proposed by Sun {\it et al.} in Ref.\ \onlinecite{sun06}. The results qualitatively agree with the calculations of the hexagonal phase\cite{kim10,silkin13} in that a VBM away from $\Gamma$ and a topological surface state is found for the Petrov-type stacking and a VBM at $\Gamma$ and no topological surface state for the KH-type stacking. The topological surface state in the Petrov phase crosses $E_\mathrm{F}$ at lower $k_\mathrm{||}$ than that of the VBM. The KH phase exhibits a VBM at $\Gamma$ for all $k_z$ while the Petrov phase shows the VBM away from $\Gamma$ for all $k_z$ in accordance with the ARPES data. The strongest surface character is found for the Rashba-type surface state at $-0.6$\,eV at $\bar{\Gamma}$ similar to the case of Sb$_2$Te$_3$.\cite{pauly12}

\bibliographystyle{apsrev}

\end{document}